\documentclass[twoside]{dis04}
\def\runauthor{L. Magnea}
\def\shorttitle{Sudakov Logs and Power Corrections}
\def\eq#1{Eq.~(\ref{#1})}

\begin{document}

\title{SUDAKOV LOGS AND POWER CORRECTIONS\\
FOR SELECTED EVENT SHAPES
}

\author{Lorenzo Magnea}

\address{Dipartimento di Fisica Teorica, Universit\`a di Torino,\\
INFN, Sezione di Torino\\
Via P. Giuria, 1, I--10125 Torino, Italy\\
{\rm and}\\
CERN, Department of Physics, Theory Division\\
CH--1211 Gen\`eve 23, Switzerland\\
E-mail: magnea@to.infn.it }

\maketitle

\abstracts{I summarize the results of two recent studies analyzing
perturbative and nonperturbative effects of soft gluon radiation on 
the distributions of the C-parameter and of the class of angularities,
by means of dressed gluon exponentiation.
}

\section{Introduction} 

Event shape distributions in high energy scattering processes such as
$e^+ e^-$ annihilation and DIS have been the focus of many theoretical
studies in recent years (for reviews, see for example
\cite{Dasgupta:2003iq} and \cite{Magnea:2002xt}). From the viewpoint
of a QCD theorist, event shape distributions are of considerable
interest because they probe the dynamics of strong interactions at a
wide range of scales, from a purely perturbative regime to
configurations dominated by soft gluon emission.  As a consequence,
the theoretical description of these distributions requires a wide
range of tools, from the computation of finite order perturbative
corrections to resummation and finally to the analysis of power
corrections.

To establish the tools required for the analysis, consider the case of
the thrust $T$. Away from the two-jet limit, $T
\to 1$, the thrust distribution is dominated by hard gluon emission
and can be computed perturbatively. Such a computation, however, is
unreliable as $\tau = 1 - T \to 0$, where the results
diverge order by order in perturbation theory, while the physical
distribution vanishes. The reason is well understood: as $\tau \to 0$
gluon radiation is forced to be soft or collinear to the primary
partons, and thus the distribution is dominated by Sudakov logarithms,
which need to be resummed in order to recover even the qualitative
features of data.

Resummation of Sudakov logarithms leads to exponentiation of the Laplace 
transform of the distribution~\cite{Catani:1992ua}. For the thrust,
\begin{eqnarray} 
      \int_0^\infty d \, \tau {\rm e}^{- \nu \tau} 
      \frac{1}{\sigma}
      \frac{d \sigma}{d \tau}~ & = & ~\exp \Bigg[ \int_0^1 \frac{d u}{u}
      \left( {\rm e}^{- u \nu} - 1 \right) \Bigg( 
      B \left(\alpha_s \left(u Q^2 \right) \right) +
      \nonumber \\ & + & ~
      \int_{u^2 Q^2}^{u Q^2}
      \frac{d q^2}{q^2} \, A \left(\alpha_s(q^2) \right) 
      \Bigg) \Bigg]~.
\label{genexp}
\end{eqnarray}
Corrections to this formula are suppressed by powers of $\nu
\Lambda/Q$, corresponding to powers of $\Lambda/(Q \tau)$ upon
inversion of the transform. The functions $A$ and $B$ are known
respectively to three and two loops, corresponding to a resummation up
to NNLL accuracy.

Although \eq{genexp} is sufficient, upon matching with finite order
results, to provide a fit of the data for values of $\tau$ larger than
those corresponding to the peak of the distribution, a complete
description requires the inclusion of power-suppressed corrections. In
fact, as $\tau$ becomes of the order of $\Lambda/Q$, all corrections
proportional to powers of $\Lambda/(Q \tau)$ must be taken into
account. Fortunately, \eq{genexp} can be used to construct a
perturbatively motivated parametrization of these
corrections~\cite{Korchemsky:1999kt}.  Introducing an IR cutoff $\mu$,
one can isolate the ambiguous contributions to \eq{genexp}, arising
from the fact that the Landau pole of the strong coupling is on the
integration contour. Power corrections thus exponentiate in the
Sudakov region, and they can be expressed in terms of the anomalous
dimension $A$, as
\begin{eqnarray}
      S_{\rm NP}(\nu/Q,\mu) & = & \int_0^{\mu^2} \frac{d q^2}{q^2}
      A \left( \alpha_s (q^2) \right)
      \int_{q^2/Q^2}^{q/Q}
      \frac{d u}{u} \left({\rm e}^{- u\nu} - 1 \right)
      \nonumber \\ & = &
      \sum_{n = 1}^\infty \frac{1}{n!} \left( \frac{\nu}{Q}
      \right)^n \lambda_n(\mu^2) + {\cal O} \left( \nu \left(
      \frac{\Lambda}{Q}\right)^2 \right)~,
\label{sudpar}
\end{eqnarray}
The moments $\lambda_n(\mu^2)$ can be organized into a ``shape
function'', which can be modeled and folded with the perturbative
distribution~\cite{Korchemsky:2000kp,Belitsky:2001ij}.

An especially compelling model for shape functions can be constructed
by combining renormalon methods with Sudakov resummations. The basic
idea is to start by computing the single gluon contribution to the
relevant cross section , with an arbitrary number of quark bubble
insertions in the gluon propagator. Integrating inclusively over the
quark pairs emitted into the final state, and summing over the number
of bubbles, this leads to the ``characteristic function'' of the
dispersive method~\cite{Ball:1995ni,Dokshitzer:1995qm}, {\it i.e.} the
mass distribution of the single virtual gluon contribution to the
desired cross section.  Under the assumption of ultraviolet
dominance~\cite{Beneke:1997sr} of power corrections, this function
encodes information about their size, and can be used to parametrize
them.  Dressed gluon exponentiation~\cite{Gardi:2001di} combines this
renormalon calculation with Sudakov resummation by using the single
dressed gluon cross section as kernel for the exponentiation, writing
\begin{equation}
      \ln \left[\left( \frac{d \tilde{\sigma}}{d \nu} \right)_{DGE} \right] =
      \int_0^\infty d \tau \left( \frac{d \sigma}{d \tau} \right)_{SDG}
      \left( 1 - e^{- \nu \tau} \right)~.
\label{kernexp}      
\end{equation}
Dressed gluon exponentiation (DGE) has several nice features. First of
all, it incorporates most of the current knowledge of the cross
section in Sudakov limit, including in particular NL logarithms to all
orders, provided the coupling is chosen appropriately. Furthermore,
since the renormalon calculation is formally exact in the
large-$n_f$ limit, all subleading logs are also included in the
same limit. One can then observe that the coefficients of formally
subleading logs grow factorially, and use this information to gauge
the range of applicability of the resummed formalism. Finally, by
imposing the constraint of energy conservation on multi-gluon emission
by means of the Laplace transform in \eq{kernexp}, DGE generates a
nontrivial pattern of exponentiated power corrections, and can be used
to construct a model for the shape function. DGE has been applied to
several high energy QCD cross sections, ranging from event
shapes~\cite{Gardi:2002bg} to inclusive DIS~\cite{Gardi:2002bk} and to
radiative and semileptonic $B$ decays~\cite{Gardi:2004ia}. In the
following we will describe the results obtained with this method in
applications to the $C$-parameter~\cite{Gardi:2003iv}, and to the
class of angularities~\cite{Berger:2004xf}.

\section{The C-parameter}

The $C$-parameter in $e^+ e^-$ annihilation has the nice feature of being 
a function of the final state momenta $p_i$ defined without reference 
to any minimization procedure, such as the one required to determine the 
thrust axis. It is well studied, both perturbatively~\cite{Catani:1998sf}
and at the level of power corrections~\cite{Dasgupta:1999mb,Smye:2001gq}.
A covariant definition is
\begin{equation}
      C = 3 - \frac32 \sum_{i, j} \frac{(p_i \cdot p_j)^2}{(p_i
      \cdot Q) \, ( p_j\cdot Q)}~.
\label{C_def}
\end{equation}
At one loop, we will need an expression for $C$ in the case of emission
of a single gluon with virtuality $\xi = k^2/Q^2$. In terms of $x_i = 2 
p_i \cdot Q/Q^2$ one finds
\begin{equation}
      {\bf c}(x_1, x_2, \xi) \, \equiv \, \frac{C}{6} \, = \,
      \frac{(1 - x_1)(1 - x_2)(1 - x_3 + 2 \xi) - \xi^2}{x_1 x_2 x_3}~.
\label{cmass}
\end{equation}
The characteristic function, corresponding to the cross section for the 
emission of a gluon with virtuality $\xi$, is then given by
\begin{equation}
    {\cal F}(\xi, c) = \int d x_1 d x_2 \, {\cal M}(x_1, x_2, \xi) \,
    \delta \left({\bf c}(x_1, x_2, \xi) - c \right)~,
\label{charf}
\end{equation}
where the one-loop matrix element for virtual gluon emission is
\begin{equation}
    {\cal M}(x_1, x_2, \xi) = \frac{(x_1 + \xi)^2 + (x_2 + \xi)^2}{(1 -
    x_1)(1 - x_2)} - \frac{\xi}{(1 - x_1)^2} - \frac{\xi}{(1 - x_2)^2}~.
\label{matrel}
\end{equation}
The characteristic function ${\cal F}(\xi, c)$ in \eq{charf} can be
computed exactly in terms of elliptic
integrals~\cite{Gardi:2003iv}. In order to perform DGE it is useful to
turn to a Borel representation of the single dressed gluon cross
section,
\begin{equation}
    \frac{1}{\sigma} \left. \frac{d \sigma}{d c}
    \right\vert_{{\rm SDG}}
    = \frac{C_F}{2 \beta_0} \, \int_0^{\infty} d u \,
    \left( Q^2/\Lambda^2 \right)^{- u} \, B(c, u)\,,
\label{sborel}
\end{equation}
where $B(c, u)$ is obtained by integrating $d {\cal F}(\xi, c)/d \xi$ over 
phase space with a weight $\xi^{-u}$. This integral cannot be performed 
exactly, but one can get an analytic answer for the terms responsible for 
Sudakov logarithms, which are singular as $c \to 0$. Having determined the 
relevant contributions to $B(c,u)$, one can exponentiate and obtain the
physical distribution by mean of an inverse Laplace transform, as
\begin{equation}
    \frac{1}{\sigma} \left. \frac{d \sigma}{d c}
    \right\vert_{{\rm DGE}} = \int_{ k - {\rm i} \infty}^{k + {\rm i} \infty}
    \frac{d \nu}{2 \pi {\rm i}} \, {\rm e}^{\nu c} \,
    \exp \left[ S\left( \nu, Q^2 \right) \right] \, , 
\label{invlapl}
\end{equation}
where
\begin{equation}
    S\left( \nu, Q^2 \right) =
    \int_0^\infty d c \, \left. \frac{1}{\sigma}
    \frac{d \sigma}{d c} \right\vert_{{\rm SDG}}
    \left({\rm e}^{- \nu c} - 1 \right) \, .
\label{expo1}
\end{equation}
The exponent admits a Borel representation
\begin{equation}
    S\left( \nu, Q^2 \right) = \frac{C_F}{2 \beta_0} \, \int_0^{\infty} d u \,
    \left(Q^2/\Lambda^2\right)^{- u} \, B_c (\nu, u) \, ,
\label{expo2}
\end{equation}
where, in the large-$n_f$ limit, one finds
\begin{eqnarray}
    B_c (\nu, u) & = & 2 \, {\rm e}^{5 u/3} \,
    \frac{\sin \pi u}{\pi u} \, \left[\Gamma(- 2 u) \left(\nu^{2 u} - 1
    \right) 2^{1 - 2 u} \frac{\sqrt{\pi} \Gamma(u)}{\Gamma(\frac{1}{2} + u)}
    \right. \nonumber \\ & - & \left.
    \Gamma(- u) \left({\nu}^{u} - 1 \right) \left(\frac{2}{u} +
    \frac{1}{1 - u} + \frac{1}{2 - u} \right) \right] \, .
\label{bc}
\end{eqnarray}
Starting from \eq{bc} one can recover perturbative Sudakov logarithms
(by expanding in powers of $u$), and one can quantify the strength of
power corrections, by looking at the location of poles in
$u$. Specifically, the second factor in \eq{bc} corresponds to
collinear radiation, and it is identical to the one found for
thrust~\cite{Gardi:2002bg}. The poles at $u = 1,2$ correspond to power
corrections of the form $\nu (\Lambda^2/Q^2)^p$, with $p = 1,2$. The
first factor, on the other hand, arises form soft radiation, and has
poles at $u = m/2$, with $m$ odd, corresponding to power corrections
of the form $\nu (\Lambda/Q)^m$. The cancellation of the pole at $u =
0$ expresses the IR-collinear safety of the $C$-parameter.

Comparing, for example, with the results for the
thrust~\cite{Gardi:2002bg}, one verifies that Sudakov logarithms are
identical for the two observables up to NLL level, as observed in
\cite{Catani:1998sf}. The pattern of power corrections is also
similar, however one finds that both the coefficients of subleading
logs and the residues of the poles corresponding to soft power
corrections are smaller for the $C$-parameter than they are for the
thrust. This can be traced back to the fact that the typical scale for
soft emissions is $2 Q c$ for the $C$-parameter, as opposed to $Q
\tau$ for the thrust. If one takes this large-$n_f$ result
seriously, one is lead to conclude that the impact of subleading
logarithms and of subleading power corrections should be smaller for
$C$ than it is for the thrust. The resummed perturbative prediction
should thus be more reliable, and the approximation of the shape
function by a shift of the perturbative distribution should work
better in this case.

\section{The class of angularities}

Angularities are a one-parameter class of event shapes introduced
in~\cite{Berger:2003iw}. They are defined by
\begin{equation}
    \tau_a = \frac{1}{Q} \sum_i (p_\perp)_i {\rm e}^{- |\eta_i| (1 - a)}~,
\label{angu}
\end{equation}
where transverse momenta and rapidities are defined with respect to
the thrust axis. For $a = 0$ one recovers the thrust ($\tau_0 = 1 - T$),
while $a = 1$ corresponds to jet broadening. Resummation of Sudakov 
logarithms was worked out in~\cite{Berger:2003iw} for $a<1$. The result
has a nontrivial $a$ dependence: for example at the LL level one finds
\begin{equation}
    \ln \left[\tilde{\sigma_{\rm LL}}
    \left(\nu, a \right)\right] =
    2 \, \int\limits_0^1 \frac{d u}{u} \Bigg[ \,
    \int\limits_{u^2 Q^2}^{u Q^2} \frac{d p_T^2}{p_T^2}
    A\left(\alpha_s(p_T)\right)
    \left( {\rm e}^{- u^{1-a} \nu \left(\frac{p_T}{Q}\right)^{a} }-1 \right)
    \Bigg]~.
\label{ll}
\end{equation}
Notwithstanding this complicated $a$ dependence, a study of power
corrections of the form $\nu (\Lambda/Q)^m$, using \eq{ll} as a
starting point, showed a remarkable scaling behavior: the shape
function suggested by the resummation for $d \sigma/d \tau_a$ depends
on $a$ only through an overall factor of $1/(1 - a)$
\cite{Berger:2003pk,Berger:2003gr}. This simple scaling arises, in the
context of resummation, from boost invariance of the eikonal cross
section responsible for logarithmic enhancements. Since DGE
complements the resummation by including the effect of subleading
logarithms in the large-$n_f$ limit, and provides an explicit model of
power corrections consistent both with the resummation and with
renormalon calculus, it was interesting to check whether the scaling
suggested in~\cite{Berger:2003pk,Berger:2003gr} would remain
valid. The test is nontrivial also because boost invariance is broken
in DGE by gluon virtuality, and it is interesting to see how it is
eventually recovered in the Sudakov limit. This study was performed
in~\cite{Berger:2004xf}.

The first step, as for the $C$-parameter, is to provide a definition of the
observable at one loop for an emitted gluon with virtuality $\xi$. The 
definition adopted in~\cite{Berger:2004xf} is
\begin{equation}
    \tau_a = \frac{(1 - x_i)^{1 - a/2}}{x_i}
    \left[(1 - x_j - \xi)^{1 - a/2} (1 - x_k + \xi)^{a/2} + 
    (j \leftrightarrow k) \right]~,
\label{defo}
\end{equation}
which has the correct limit as $\xi \to 0$ and is simple enough to allow 
for analytic computations. It can be shown that leading power corrections
are not affected by changes of \eq{defo} which are analytic in $\xi$.

In this case, it is not possible to compute the characteristic
function in closed form. A detailed study of the limit of soft
radiation leads anyhow to a simple expression for the soft
contribution to the Borel function of DGE, corresponding to the first
erm of \eq{bc}. One finds
\begin{equation}
    B_a^{\rm soft} (\nu, u) = \frac{1}{1 - a} \left[ 2 \,
    {\rm e}^{5 u/3} \, \frac{\sin \pi u}{\pi u} \, \Gamma(- 2 u)
    \left(\nu^{2 u} - 1 \right) \frac{2}{u} \right]~,
\label{scale}
\end{equation}
exactly the scaling behavior predicted by the resummation. DGE also
provides a model for power corrections of collinear origin. Although
these in principle may be affected by the choice of the massive
definition of the observable, it is interesting to notice that they
are suppressed by a power of $Q$ which grows as $a$ becomes large and
negative. One finds that collinear power corrections
are suppressed at least by $\nu (\Lambda/Q)^{2 - a}$. Comparing the thrust
distribution to the angularity distribution for a negative value of
$a$ should thus provide a simple and clean test of the scaling rule,
largely unaffected by errors due to subleading power
corrections.

\section{Perspective}

Studies of event shape distributions in and beyond perturbative QCD
have reached a considerable degree of refinement, and provide robust
theoretical predictions which in some cases should be fairly easy to
test experimentally. Such tests are indeed desirable, because they
would strongly constrain our current understanding of the transition
between perturbative and nonperturbative QCD, and they might have
practical consequences, for example on current determinations
$\alpha_s$~\cite{Jones:2003yv}.  The fact that some of the recent
theoretical progress has taken place as the work of the LEP
collaborations was winding down is a warning for the future: ``old''
data may well contain a wealth of unexplored information, so it should
continue to be possible to perform new analyses, as theory progresses
or new viewpoints emerge.

\section*{Acknowledgements} I thank my collaborators, Carola Berger and
Einan Gardi, for their essential contribution to the results described 
in this paper.

\end{document}